\documentstyle[prl,twocolumn,aps,floats,epsfig]{revtex}

\begin{document}

\draft
\twocolumn[\hsize\textwidth\columnwidth\hsize\csname
@twocolumnfalse\endcsname

\title{Evidence for a Quasi-1D Topological-Excitation 
Liquid in Bi$_{2}$Sr$_{2}$CaCu$_{2}$O$_{8+x}$ from Tunneling 
Spectroscopy}
\author{A. Mourachkine}
\address{Universit\'{e} Libre de Bruxelles, CP-232, 
Blvd du Triomphe, B-1050 Brussels, Belgium}

\date{received 26 March 2001; {\bf Europhysics Lett.}} 

\maketitle

\begin{abstract}
Tunneling measurements have been carried out on heavily 
underdoped and slightly overdoped Bi$_{2}$Sr$_{2}$CaCu$_{2}$O$_{8+x}$ 
(Bi2212) single crystals by using a break-junction technique. We find 
that {\em in}\,-plane tunneling spectra below $T_{c}$ are the 
combination of incoherent part from the pseudogap and coherent 
quasiparticle peaks. There is a correlation between the 
magnitude of the pseudogap and the magnitude of the 
superconducting gap in Bi2212. We find that the quasiparticle 
conductance peaks are caused by condensed solitonlike excitations. 
\end{abstract}

]

Recent intrinsic $c$\,-axis tunneling data obtained in 
Bi$_{2}$Sr$_{2}$CaCu$_{2}$O$_{8+x}$ (Bi2212) mesas 
show that the pseudogap (PG) is a normal-state gap, and the PG
and the superconducting gap (SG) coexist below $T_{c}$ 
\cite{Yurgens}. Thus, the PG in Bi2212 arises either from charge-density 
waves (CDW) or from local antiferromagnetic (AF) correlations [or 
spin-density waves (SDW)] \cite{Yurgens,Mark,Klemm}. There is a consensus 
on doping dependence of the PG in cuprates: the magnitude of the PG 
decreases with increase of hole concentration \cite{Mark,Klemm,Timusk}.

Figure 1(a) shows a theoretical $I(V)$ tunneling characteristic in a 
superconductor-insulator-normal metal (SIN) junction (fig. 6 in 
ref.\cite{Tinkham}). In the {\em tunneling} regime, it is expected 
that the $I(V)$ curve at high positive (low negative) bias, 
depending on the normal resistance of the junction, lies somewhat below 
(above) the normal-state curve.
In conventional superconductors (SCs), the Blonder-Tinkham-Klapwijk 
(BTK) predictions are verified by tunneling experiments \cite{Wolf}. 
However, in cuprates, the BTK theory is violated. 
Figure 1(b) shows the SC-insulator-SC (SIS) $I(V)$ curve measured in an 
underdoped Bi2212 single crystal with $T_c$ = 83 K (fig. 1 in 
ref.\cite{Miyakawa}). In fig. 1(b), one can see that the $I(V)$ 
curve at high positive (low negative) bias passes not below 
(above) the straight dash line \cite{note} but far above (below) the 
line. This fact cannot be explained by the d-wave symmetry of 
the order parameter. To our 
knowledge, this question has never been raised in the literature 
before. This finding is the main motivation of the present work.

To the best of our knowledge, the soliton SC was for the first time 
considered in ref.\cite{bisoliton} in order to explain the SC in organic 
quasi-one-dimensional (quasi-1D) conductors. Later, Davydov 
\cite{Davydov1,Davydov2} applied the model of soliton SC to cuprates. 
The theory  is based on the concept of {\em bisolitons}, or electron (or hole) 
pairs coupled in a singlet state due to local deformation of the lattice.

Tunneling spectroscopy is an unique probe of SC state in that it can, in 
principle, reveal the quasiparticle (QP) excitation density 
of states (DOS) directly with high energy resolution.
In this paper, we present tunneling measurements performed on 
heavily underdoped and slightly overdoped
Bi2212 single crystals by using a break-junction technique. We find 
that {\em in}\,-plane tunneling spectra below $T_{c}$ are the 
combination of incoherent part from the PG and coherent QP 
peaks. There is a correlation between the magnitude of the 
PG and the magnitude of the SG in Bi2212. We find that the QP 
conductance peaks are caused by condensed topological solitons. Thus, we 
present an evidence for a quasi-1D topological-excitation liquid in Bi2212.

The overdoped Bi2212 single crystals were grown using a self-flux 
method as described elsewhere \cite{AMour1}.
The underdoped samples were obtained from the overdoped single 
crystals by annealing them in vacuum. The $T_{c}$ value was determined 
by the four-contact method. The transition width is less than 1 K in
\begin{figure}[t]
\leftskip-10pt
\epsfxsize=1.0\columnwidth
\centerline{\epsffile{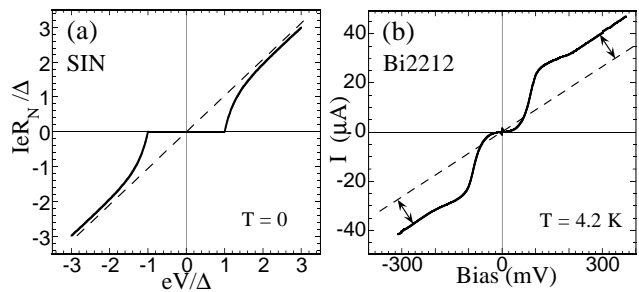}}
\vspace{2mm}
\caption{(a) Theoretical $I(V)$ tunneling characteristic in a SIN 
junction of a SC with the isotropic energy gap \protect\cite{Tinkham}.
The dash line shows the normal-state curve. 
(b) Measured $I(V)$ curve in a SIS junction of an underdoped Bi2212 
with $T_{c}$ = 83 K \protect\cite{Miyakawa}. The dash 
line which is parallel to the $I(V)$ curve at high bias is a guide 
to the eye. The arrows show the offset from the dash line. One can 
immediately notice the difference between the two plots 
\protect\cite{note}.}
\label{fig1}
\end{figure} 
the overdoped crystals, and a few degrees in the underdoped Bi2212.  

Experimental details of our break-junction setup can be found 
elsewhere \cite{Hancotte}. In short, many break-junctions were 
prepared by gluing a sample with epoxy on a flexible insulating 
substrate, and then were broken in the $ab$\,-plane by bending the
substrate with a  differential screw at low temperature in a He
ambient. The electrical contacts (typically with 
the resistance of a few ohms) are made by attaching gold wires 
to a crystal with silver paint. The $I(V)$ and $dI/dV(V)$ 
characteristics are determined by the four-terminal 
method by using a standard lock-in modulation technique.
At low (constant) temperature, in one junction, we usually obtain a 
few tunneling spectra by changing the distance between broken parts of 
a crystal, going back and forth {\em etc}., and, every time, the tunneling
occurs most likely in different places.

Figure 2(a) shows the SIS $dI/dV(V)$ and $I(V)$
obtained in an underdoped Bi2212 single crystal,  
which look like usual spectra in Bi2212 \cite{Miyakawa2}. 
In fig. 2(a), the Josephson $I_{c}R_{n}$ product is estimated to be 
13.4 mV. The gap magnitude, $\Delta_{sc}$= 64 meV, is in 
good agreement with other tunneling measurements \cite{Miyakawa2}. 
The $dI/dV(V)$ and $I(V)$ shown in fig. 2(b) are
\begin{figure}[t]
\leftskip-10pt
\epsfxsize=1.0\columnwidth
\centerline{\epsffile{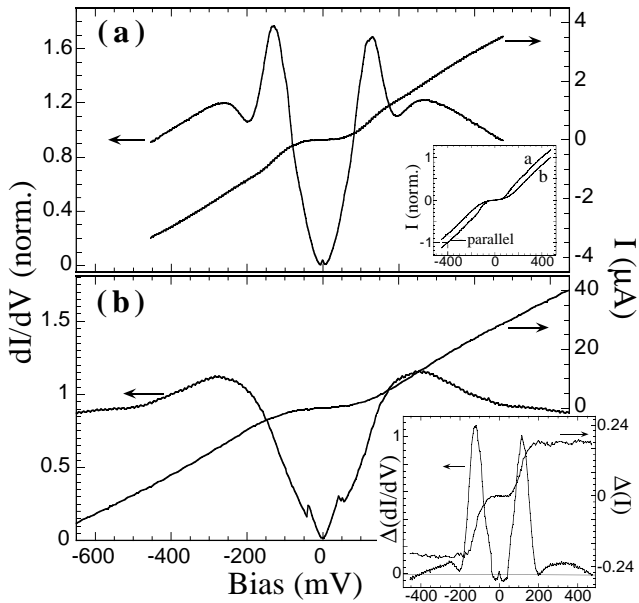}}
\vspace{2mm}
\caption{SIS $dI/dV(V)$ 
and $I(V)$ measured at 14 K 
within the same underdoped Bi2212 single crystal with $T_{c}$ = 51 K.
The $dI/dV(V)$ in both plots are normalized at -400 mV. The inset 
in the plot (b) shows the differences $(dI/dV)_{a}- (dI/dV)_{b}$ 
and $I_{a,norm}- I_{b,norm}$. The inset in the plot (a) shows how 
$I_{a}$ and $I_{b}$ are normalized: $I_{a}$ is normalized at 
-400 mV, and $I_{b}$ is adjusted to be parallel at high bias to 
$I_{a}$ (such procedure is equivalent to the normalization at 
$\pm\infty$).}
\label{fig2}
\end{figure}  
obtained within the {\em same} underdoped single crystal as those in 
fig. 2(a). In fig. 2(b), the gap having $\Delta$ = 130 meV is too large to 
be a SG. It is suggestive that the spectra in 
fig. 2(b) correspond to the PG. The SC in Bi2212 is weak in the heavily 
underdoped region \cite{Tallon,Shen}. This may explain why it 
is possible to observe separately the PG in the heavily underdoped 
Bi2212 by taking into account that tunneling spectroscopy probes the 
local DOS. The absence of the Josephson current in the spectra shown 
in fig. 2(b) indicates that the humps in the conductance are incoherent.
The differences between the spectra 
shown in figs 2(a) and 2(b) are presented in the inset of fig. 2(b), 
which correspond to a "pure SG". Some parts of the $dI/dV(V)$ in the 
inset of fig. 2(b) are slightly below zero because the spectra
in figs 2(a) and 2(b) are not taken under the exact same conditions.
The small humps in the $dI/dV(V)$ shown in the inset of fig. 2(b) are 
discussed below. The $dI/dV(V)$ and $I(V)$ curves in the inset of 
fig. 2(b) resemble the characteristics of a bound state of two solitons 
(a bi-soliton), shown in the inset of fig. 3(b). 
Thus, we find that the conductance in fig. 2(a) consists 
of the two contributions: the humps which correspond to the PG (which is 
a normal-state gap \cite{Yurgens,Mark,Klemm}), and the QP peaks which 
are caused by condensed solitonlike excitations. In fig. 2(b), one can see 
that the PG is anisotropic.

Figures 3(a) and 3(b) show the SIS $dI/dV(V)$ and $I(V)$ 
\begin{figure}[t]
\leftskip-10pt
\epsfxsize=1.0\columnwidth
\centerline{\epsffile{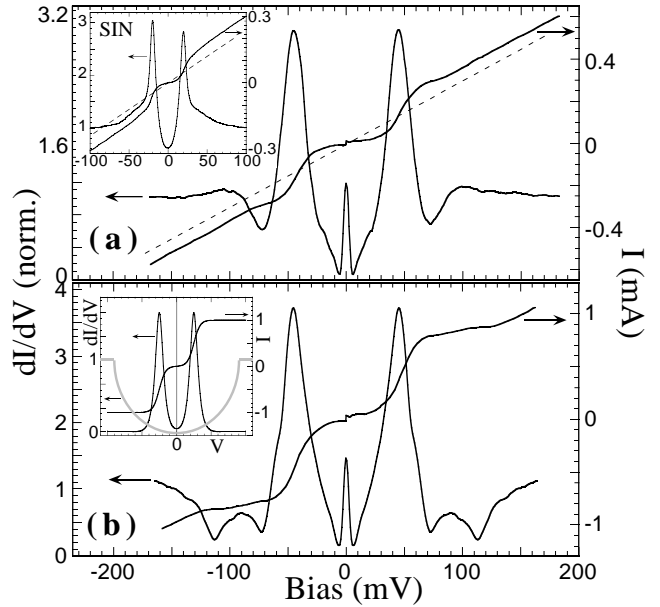}}
\vspace{2mm}
\caption{SIS $dI/dV(V)$ and $I(V)$
measured at 14 K within the same overdoped Bi2212 single crystal 
with $T_{c}$ = 88 K. The inset in the plot (a) shows SIN $dI/dV(V)$ and 
$I(V)$ measured at 9 K in an overdoped Bi2212 with 
$T_{c}$ = 87.5 K (the same axis parameters as main plot).
The dash lines in the plot (a) and in the inset, which
are parallel at high bias to the $I(V)$ curves, are guides to the eye. The 
inset in the plot (b) shows the normalized characteristics of a bound 
state of two solitons (a bisoliton) \protect\cite{french,Mostovoy}. The gap 
is shown {\em schematically} in grey (the $I(V)$ of the gap is not shown). 
The hight of conductance peaks depends on the density of added 
(removed) electrons \protect\cite{Mele}.}
\label{fig3}
\end{figure} 
obtained within the {\em same} overdoped Bi2212 single 
crystal. In figs 3(a) and 3(b), the Josephson $I_{c}R_{n}$ product
is estimated to be 6 and 7.5 mV, respectively. The characteristics in fig. 3(a) 
look like usual spectra in Bi2212 \cite{Miyakawa2}. The $dI/dV(V)$ 
and $I(V)$ curves in fig. 3(b) resemble the spectra in the inset of fig. 2(b), 
and the bisoliton characteristics shown in the inset of fig. 3(b). This means 
that the contribution from the PG in the spectra shown in fig. 3(b) is small 
in comparison with the contribution from the QP peaks, at least, at low bias. 
This is most likely due to the fact that, in slightly overdoped cuprates, the 
SC is the strongest, and the "strength" of the PG is weak 
\cite{Tallon,Shen}. At high bias, the contribution from the PG will be
always predominant, even, if the PG is weak.

We also performed measurements in Ni-doped Bi2212 single crystals 
(overdoped in oxygen): the measured data which are presented in an 
extended paper \cite{AMour9} are similar to the data in the overdoped 
Bi2212, shown in fig. 3. From the temperature dependence of $dI/dV(V)$, 
it is clear that the small humps which appear in $dI/dV(V)$ 
at bias twice as large as the bias of the QP peaks, shown in fig. 3(b), 
relate to the QP peaks and not to the PG. 
The humps are also observed in the Ni-doped Bi2212. In the inset 
of fig. 2(b), a similar hump is present in the conductance at 
negative bias. These humps can be well understood in terms of a 
{\em nanopteron} soliton \cite{french} which is discussed below.

Since the spectra measured in an underdoped Bi2212, shown in the inset 
of fig. 2(b), and the spectra measured in an overdoped Bi2212 (and
Ni-doped Bi2212), shown in fig. 3(b), are similar, the  data obtained
in underdoped and overdoped Bi2212 are consistent with each other.

In order to be sure that we observe not a SIS-junction effect 
but an intrinsic effect, we performed measurements in the
overdoped Bi2212 crystals by SIN junctions. Pt-Ir wires 
sharpened mechanically are used as normal tips.
The inset of fig. 3(a) shows the SIN $dI/dV(V)$ and $I(V)$ 
obtained in an overdoped Bi2212. In fig. 3(a), one can
see that, basically, there is no difference between the $I(V)$
characteristics measured in SIS and SIN junctions [see the dash lines
in fig. 3(a) and in the inset of fig. 3(a)].

First, we give a description of a topological soliton. The topological soliton 
is an extremely stable nonlinear excitation which can be moving or entirely 
static \cite{french,german,Mostovoy}. The solitons have particlelike 
properties, and, in solids, they occupy the intragap states in a CDW or SDW 
gap. The inset of fig. 3(b) shows the bisoliton characteristics. Since we 
consider a general solution, the gap shown {\em schematically} in the inset 
of fig. 3(b) is either a CDW or SDW gap. 

We now compare the measured data with theory. Soliton and bisoliton 
characteristics are described by hyperbolic functions 
\cite{french,german,Mostovoy}. Since the bisoliton conductance peaks 
shown in the inset of fig. 3(b) look very similar to the conductance peaks 
not only of high-$T_{c}$ SCs but also of low-$T_{c}$ SCs (not the 
background), we rely here exclusively on the $I(V)$ characteristics 
which are {\em conceptually} different for the two models: the BTK model 
for 3D case and the model based on the concept of quasi-1D topological 
excitations.

Figure 4(a) shows the measured $I(V)$ curve 
from the inset of fig. 2(b). In fig. 4, for simplicity, we 
analyze the data only at positive bias. As shown in fig. 4(a), the data 
from the inset of fig. 2(b) can be fitted very well by the hyperbolic 
function $f(V)$ = $A$$\times$$(tanh[(eV$ - 2$\Delta$$)/eV_{0}] 
+ tanh[(eV$ + 2$\Delta$$)/eV_{0}])$, where $e$ is the electron 
charge; $V$ is the bias; $\Delta$ is the maximum SC energy gap, and 
$A$ and $V_{0}$ are the constants. In fig. 4(a), we also present the 
measured $I(V)$ of the PG from fig. 2(b). We find that any $I(V)$ 
characteristic obtained in Bi2212 can be resolved 
into the two components shown in fig. 4(a): from the quasi-1D 
topological excitations and from the PG. The "usual" $I(V)$ and $dI/dV(V)$ 
spectra in Bi2212 show the presence of both components [see figs 2(a) and 
3(a)]. The absence [see fig. 2(b)] or weak contribution of one component
\begin{figure}[t]
\leftskip-10pt
\epsfxsize=1.0\columnwidth
\centerline{\epsffile{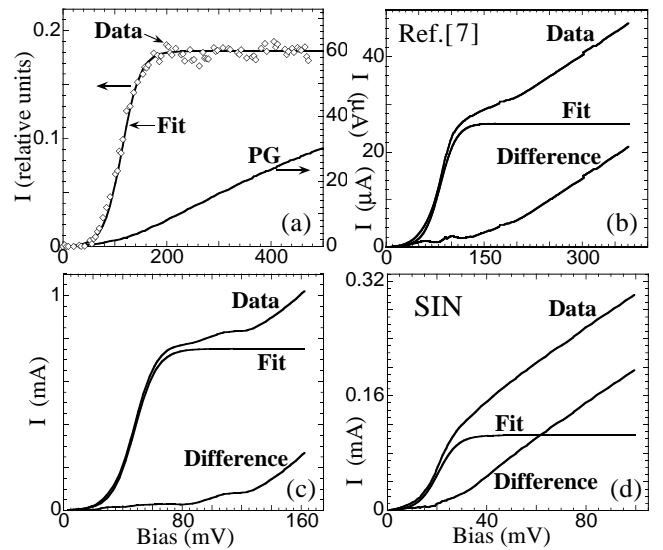}}
\vspace{2mm}
\caption{Measured $I(V)$ curves and the $f(V)$ fit (see text): 
(a) The measured data (diamonds) from the inset of fig. 2(b), the 
$I(V)$ of the PG from fig. 2(b), and the $f(V)$ 
fit. (b) The data from fig. 1(b), the $f(V)$ fit, and their difference. 
(c) The $I(V)$ from fig. 3(b), the $f(V)$ fit, and their 
difference. (d) The SIN $I(V)$ from the inset of 
fig. 3(a), the fit (see text) and their 
difference. The Josephson currents are removed.}
\label{fig4}
\end{figure}  
[see fig. 3(b)] in spectra makes the appearance of the spectra "unusual".

Figure 4(b) shows the data from ref.\cite{Miyakawa}, the $f(V)$ fit, 
and their difference. Figure 4(c) depicts the two components in the 
$I(V)$ curve from fig. 3(b). As shown in fig. 4(d), 
the contribution from the QP peaks in the SIN $I(V)$ from the inset 
of fig. 3(a) seems to be weaker than that in SIS junctions. As seen in 
fig. 4, all plots are similar. To fit the SIN $I(V)$, we use the same 
$f(V)$ function by substituting 2$\Delta$ for $\Delta$ \cite{note2}. 
In figs 4(b)--4(d), the amplitude, $A$, 
of the $f(V)$ fit can be changed, this only affects the scale but not 
the shape of the differences which correspond to the PG. The $I(V)$ curves 
in fig. 3(a) and in refs \cite{Miyakawa} and \cite{Miyakawa2} can be 
resolved into the two components in the same manner.
So, we conclude that, in Bi2212, the $I(V)$ characteristics 
of the QP peaks {\em definitely} disagree with the BTK
theory, and are in good agreement with the theory of quasi-1D 
topological excitations 
\cite{Davydov1,Davydov2,french,german,Mostovoy}.

The coherent QP peaks in $dI/dV(V)$ curves can be fitted very well by the 
derivative $[f(V)]'$ = $A_{1}$$\times$$((sech[(eV$ - 
2$\Delta$$)/eV_{0}])^{2} + (sech[(eV$ + 2$\Delta$$)/eV_{0}])^{2})$ 
\cite{AMour9}. The fit is applicable only to the QP peaks, and not to the 
humps at high bias, which correspond to the PG.

In figs 2(a) and 3(a), the ratio of hump bias to QP-peak bias in the
conductance remains more or less constant with doping, 
2 (underdoped)--2.3 (overdoped), which is in good agreement with other 
tunneling data \cite{Miyakawa2}. Since the humps correspond to the PG,
the magnitudes of the PG and SG in Bi2212 correlate with each other.

We now focus on the small humps which appear in $dI/dV(V)$ outside
the gap structure [see, {\em e.g.}, fig. 3(b)]. They can be well understood in 
terms of a nanopteron soliton which is a bound state resulting from the 
nonlinear interaction between the soliton and the periodic wave 
(see fig. 6.27(c) in ref.\cite{french}). Instead of flat asymptotics, the 
characteristics of a nanopteron have oscillations. The bisoliton can also 
exist in resonance with small amplitude linear waves \cite{Buryak}. 
The oscillations in asymptotics can be also seen in the $I(V)$ shown in 
figs 4(a) and 4(c). What kind of periodic waves in Bi2212 can interact with 
the topological excitations and cause these oscillations? Phonons are the 
primary candidate on this role. On the other hand, since the magnetic 
resonance peak observed in inelastic neutron scattering spectra 
\cite{Mignod} is caused by a magnon-like excitation (spin 1, charge 0), 
it is most likely that this spin excitation is in resonance with the 
bound state of topological solitons.

It is well known that $dI/dV(V)$ obtained in SIS junctions 
often have subgap structures [see, {\em e.g.}, fig. 3(a)]. These subgap 
structures can be also understood in terms of a nanopteron soliton. They 
have most likely the same origin as the oscillations at high bias, {\em i.e.} 
due to the nonlinear interaction between the bisoliton and 
the periodic wave (see fig. 4 in ref.\cite{Buryak}).

We now discuss the PG. In general, quasi-1D topological excitations reside 
on quasi-1D "objects". Recently, quasi-1D charge stripes separated by 2D 
AF domains have
been discovered in some cuprates \cite{Tranquada,Mochalk,Lavrov,AMour2}. 
The presence of soliton excitations on stripes implies that the stripes 
are {\em insulating}, as predicted \cite{Zaanen}, and not conducting
as widely believed. In other words, there is a CDW gap along stripes.
Then, in order to pull out an electron (a hole) from a stripe, it is necessary, 
first of all, to overcome the CDW gap on the stripe. Consequently, the 
{\em tunneling} PG is predominantly the CDW gap. The same conclusion 
can be independently obtained from the analysis of the data: The 
magnitude of PG, $\Delta_{ps}$= 130 meV, shown in fig. 2(b), is 
too large to be explained by the development of local AF correlations, 
since the value of the superexchange energy, $J \simeq$ 120 meV 
\cite{Bourge}, is not large enough to fit the data. Moreover, the magnitude 
of PG becomes larger than 130 meV at lower doping. Consequently, the 
{\em tunneling} PG is most likely a CDW gap. 

It is important to emphasize that the PG in {\em transport}
measurements \cite{Wuts,Tallon} is {\em different} from the tunneling PG.
In transport measurements, the PG relates to a spin gap into local AF 
domains \cite{Mochalk}. As an example, if a molecular chain is embedded 
into a medium, a frictional force acts on the soliton \cite{Davydov2}.

We now turn to the model of SC in cuprates. 
The ''flat'' (constant) asymptotics of $I(V)$ characteristics are the 
fingerprints of one-dimensionality. To the best of our 
knowledge, there are only two theoretical models of SC in cuprates which 
are based on the presence of one-dimensionality: the bisoliton model 
\cite{Davydov1,Davydov2} and the stripe model \cite{Emery}. In the stripe 
model, charge stripes are assumed to be metallic \cite{Emery}, thus 
contrary to the experiment. By contrast, the bisoliton model is 
{\em quantitatively} in good agreement with the data: the bisoliton model 
predicts that by increasing the hole concentration the magnitude of the 
pairing gap decreases (see fig. 3 in ref.\cite{Miyakawa}). However, the 
bisoliton model is a theory of soliton pairing, but the mechanism 
of the establishment of phase coherence is not considered. Experimentally, 
spin fluctuations mediate the phase coherence in cuprates \cite{AMour4}. 
The bisolitons are formed due to electron-phonon interactions which are 
moderately strong and nonlinear \cite{Davydov1,Davydov2}.

Lastly, it is worth noting that the analysis of many data measured in 
cuprates shows that the data can be naturally understood in the framework 
of the quasi-1D topological-excitation-liquid scenario \cite{AMour9}.

In summary, tunneling measurements have been carried out on 
underdoped and overdoped Bi2212 single crystals. 
Tunneling spectra below $T_{c}$ are the combination of 
incoherent part from the pseudogap and coherent 
quasiparticle peaks. There is a correlation between the 
maximum magnitude of the pseudogap and the distance between the 
quasiparticle peaks. The $I(V)$ characteristics of the
quasiparticle peaks are in good agreement with the theory of quasi-1D 
topological excitations. The solitons and bisolitons reside most likely on 
charge stripes. It seems that magnon-like excitations which cause the 
appearance of the magnetic resonance peak in inelastic neutron scattering 
spectra are in resonance with the bisolitons. The bisoliton model of 
superconductivity in cuprates \cite{Davydov1,Davydov2} is correct in the 
description of pairing characteristics, but it lacks the mechanism of the 
establishment of phase coherence.

I would like to thank D. Davydov, A. V. Buryak, M. V. Mostovoy and K. Maki 
for discussions, and N. Miyakawa for sending the data from 
ref.\cite{Miyakawa}.


\begin{references}
\bibitem{note} The dash line in fig. 1(b) is not the normal-state 
curve. However, any straight line passing through zero will show that 
the behavior of the $I(V)$ curve in fig. 1(b) deviates from 
the theory \cite{Tinkham}.
\end{references}
\end{document}